# Topological colloids

Bohdan Senyuk[1,2], Qingkun Liu[1,3], Sailing He[3], Randall D. Kamien[4], Robert B. Kusner[5], Tom C. Lubensky[4] and Ivan I. Smalyukh[1,2,6,*]

[1]Department of Physics, University of Colorado, Boulder, Colorado 80309, USA.

[2]Department of Electrical, Computer, and Energy Engineering, Materials Science and Engineering Program, and Liquid Crystal Materials Research Center, University of Colorado, Boulder, Colorado 0309, USA.

[3]Centre for Optical and Electromagnetic Research, Zhejiang University, Hangzhou 310058, China.

[4]Department of Physics and Astronomy, University of Pennsylvania, Philadelphia, Pennsylvania 19104, USA.

[5]Department of Mathematics and Statistics, University of Massachusetts, Amherst, Massachusetts 01003, USA.

[6]Renewable and Sustainable Energy Institute, National Renewable Energy Laboratory and University of Colorado, Boulder, Colorado 80309, USA.

*Email: ivan.smalyukh@colorado.edu

Smoke, fog, jelly, paints, milk and shaving cream are common everyday examples of colloids[1], a type of soft matter consisting of tiny particles dispersed in chemically distinct host media. Being abundant in nature, colloids also find increasingly important applications in science and technology, ranging from direct probing of kinetics in crystals and glasses[2] to fabrication of third-generation quantum-dot solar cells[3]. Because naturally occurring colloids have a shape that is typically determined by minimization of interfacial tension (for example, during phase separation) or faceted crystal growth[1], their surfaces tend to have minimum-area spherical or topologically equivalent shapes such as prisms and irregular grains (all continuously deformable—homeomorphic—to spheres). Although toroidal DNA condensates and vesicles with different numbers of handles can exist[4–7] and soft matter defects can be shaped as rings[8] and knots[9], the role of particle topology in colloidal systems remains unexplored. Here we fabricate and study colloidal particles with different numbers of handles and genus $g$ ranging from 1 to 5. When introduced into a nematic liquid crystal—a fluid made of rod-like molecules that spontaneously align along the so-called "director"[10]—these particles induce three-dimensional director fields and topological defects dictated by colloidal topology. Whereas electric fields, photothermal melting and laser tweezing cause transformations between configurations of particle-induced structures, three-dimensional nonlinear optical imaging reveals that topological charge is conserved and that the total charge of particle-induced defects always obeys predictions of the Gauss–Bonnet and Poincaré–Hopf index theorems[11–13]. This allows us to establish and experimentally test the procedure for assignment and summation of topological charges in three-dimensional director fields. Our findings lay the groundwork for new applications of colloids and liquid crystals that range from topological memory devices[14], through new types of self-assembly[15–23], to the experimental study of low-dimensional topology[6,7,11–13].

Although a coffee mug and a doughnut look different to most of us, they are topologically equivalent solid tori or handlebodies of genus $g = 1$, both being different from, say, balls and solid cylinders of genus $g = 0$, to which they cannot be smoothly morphed without cutting[11,12]. In a similar way, molecules can form topologically distinct structures including rings, knots and other molecular configurations satisfying the constraints imposed by chemical bonds[24]. Although the topology of shapes, fields and defects is important in many phenomena and in theories ranging from the nature of elementary particles to early-Universe cosmology[25,26], topological aspects of colloidal systems (composed of particles larger than molecules and atoms but much smaller than the objects that we encounter in our everyday life) are rarely explored. Typically dealing with particles with surfaces homeomorphic to spheres, recent studies[8,9,15,18,20–23] demonstrate that the topology of curved surfaces dictates the formation of defects during two-dimensional colloidal crystallization at fluid interfaces as well as inside liquid crystal droplets and around spherical inclusions in liquid crystals. However, despite the fact that several techniques for scalable fabrication of particles with complex geometric shapes and $g > 0$ have recently been introduced[18,19,27–29], the potential impact of particle topology on colloidal alignment, self-assembly and response to fields remains unexplored.

To study the interplay of particle topology and defects in liquid crystals, we fabricated topologically distinct silica particles with planar symmetry and handlebody topology of genus $g$ varying from 1 to 5; their surfaces had an Euler characteristic $\chi = 2 - 2g$ ranging from 0 to $-8$ (Fig. 1 and Supplementary Fig. 1). These particles had 1 μm × 1 μm rounded square cross-sections and ring diameters ranging from 5 to 10 μm. Handlebody particles were introduced into a nematic liquid crystal, pentyl cyanobiphenyl, and the ensuing dispersion was infiltrated into cells bounded by parallel uniformly separated substrates treated to impose either perpendicular (homeotropic) or parallel alignment of the director $n$ and thereby to create a uniform director $n_0$ in their interior in the absence of inclusions. Before dispersion, the surfaces of particles were also treated to induce perpendicular boundary conditions for $n$. The director field $n(r)$ around these handlebody colloids, which approaches $n_0$ at large distances, was probed optically by a combination of transmission-mode polarizing microscopy (PM) and three-photon excitation fluorescence polarizing microscopy (3PEF-PM)[30], schematically shown in Supplementary Fig. 2. Holographic optical tweezers allowed non-contact optical manipulation of particles at laser powers of 5–50 mW and local photothermal melting of the liquid crystal into an isotropic state at powers of about 100 mW and higher. Because of the strong surface anchoring and rounded cross-section of the colloids, quenching the liquid crystal from the isotropic to the nematic phase creates director configurations that vary smoothly away from homeotropic alignment at the particle surfaces and that also exhibit bulk defects.

Colloidal handlebodies spontaneously align with their ring planes either perpendicular or parallel to $n_0$ (Figs 1 and 2 and Supplementary Figs 3–9). The prevailing alignment of handlebodies perpendicular to $n_0$ is more common because it minimizes the elastic free energy of $n(r)$ distortions induced by the particles with perpendicular boundary conditions. Handlebodies aligned with ring planes parallel to $n_0$ are obtained by melting and subsequently quenching the surrounding liquid crystal with laser tweezers (Fig. 2). These particles can also be made to align parallel to or obliquely to $n_0$ by confinement in cells of thickness comparable to their lateral dimensions (Supplementary Fig. 10). Handlebodies are elastically repelled from both confining substrates as a result of strong surface anchoring conditions. However, because of the density mismatch between silica and the liquid crystal, they tend to rest somewhat below the cell midplane, where gravity is balanced by the elastic forces (Supplementary Fig. 9).

Optical micrographs obtained by using different imaging modalities (Fig. 1a–h) reveal that handlebody colloids aligned perpendicular to $n_0$ are all surrounded by single half-integer exterior disclination loops of topological point defect (hedgehog) charge $m = -1$ but have different defects within their interiors (Fig. 1i–l). Each genus-$g$ particle has $g$ defects in its holes, which are either singular disclination loops or hyperbolic point defects of topological hedgehog charge $m = +1$. Disclination loops in the holes of each handlebody can be transformed into point defects and vice versa by melting the liquid crystal into an isotropic state with tweezers of laser power more than 100 mW and subsequently quenching into a nematic phase, indicating that free energies due to director configurations with these defects are comparable. These hedgehog charges of the point defects and disclination loops have been determined by assuming that the vector field lines point perpendicularly outwards from the particle surfaces (Fig. 3) and by mapping the vector fields around particles, point defects and disclination loops onto the order-parameter space[31,32]. Because $n$ has nonpolar symmetry (that is, $n$ is equivalent to $-n$), one could have chosen the vector field pointing inwards to the surface of colloids, which would consequently reverse the signs of all hedgehog charges induced by particles in a uniformly aligned liquid crystal. The relative charges of all the defects would remain the same, as would the net charge of 0, ensuring conservation of topological charge.

Colloidal particles oriented with their rings parallel to $n_0$ tend to induce point defects both within the holes and next to the particles (Fig. 2). The point defects occasionally open into disclination loops that follow the curved edge faces of particles and have a topological charge equivalent to that of the point defect that they replace (Fig. 2b, e, h, k). Although the handlebodies oriented perpendicular and parallel to $n_0$ induce a different director field $n(r)$, the sum of hedgehog charges due to induced point defects and disclination loops, $\Sigma_i m_i = -m_c = \pm\chi/2$, compensates for the colloidal particles' hedgehog charge $m_c$ due to $n(r)$ at their surfaces and is uniquely predetermined by particle topology (Fig. 3). The signs depend solely on the choice of the direction of the vector field at the surface of particles. This relation holds for all colloidal handlebodies ($g = 1, 2,…, 5$) and for spherical colloids with $g = 0$ and $\Sigma_i m_i = \pm\chi/2 = \pm 1$ (Supplementary Fig. 12) studied previously[15], and can be understood using simple considerations based on the Gauss–Bonnet theorem[12]. Recall that the topological charge $m_c$ of any region of space V bounded by a surface $S = \partial V$ is the degree[11–13] of $n$ along S, which can be calculated by integrating the Jacobian of $n(r)$ over that surface[10,32,33], $m_c = (1/4\pi)\int_S dx_1 dx_2 \, n \cdot \partial_1 n \times \partial_2 n$. Because $n(r)$ aligns with the (outer) unit normal field to the colloidal surface S, the integral reduces to the total Gauss curvature of S divided by $4\pi$. The Gauss–Bonnet theorem[12] states that the total Gauss curvature of a closed surface without boundary is quantized in units of $4\pi$ equal to $4\pi(1 - 2g) = 2\pi\chi$ and remains unchanged during all continuous deformations of the surface; it follows that the hedgehog charge $m_c$ of $n(r)$ along S is (up to sign) $m_c = \pm 2\pi\chi/(4\pi) = \pm(1 - g)$. Because the director is roughly constant (along $n_0$) far from the colloidal inclusions, an imaginary surface surrounding the colloids and all other defects will have a net zero charge. It follows that the sum of the defect charges must cancel the degree on the colloidal surface S, and so the total hedgehog charge of point defects and disclination loops will be $\Sigma_i m_i = -m_c = \pm\chi/2 = \pm(1 - g)$, regardless of the orientation of the particles with respect to $n_0$, as observed experimentally.

The diagram in Fig. 3g shows that both interior and exterior disclination loops of the configurations shown in Fig. 1 can be transformed to hyperbolic point defects of equivalent hedgehog charge. Furthermore, these structures can be also transformed into a nonsingular twist-escaped looped $n(r)$ configuration with a net topological hedgehog charge equal to zero (Fig. 3h) and resembling the "bubble gum" structure studied previously[21,22]. Although perpendicular boundary conditions due to the handlebody-shaped particles in the liquid crystal with a uniform $n_0$ can be satisfied by a minimum number of point or ring defects of the same sign having the total hedgehog charge of $\pm\chi/2$ (that is, no singularities for a solid torus, as shown in Fig. 3h, and $g - 1$ point or ring defects for a handlebody of genus $g$), these field configurations relax to topology-satisfying $n(r)$ that also minimize the free energy. The energetic cost of introducing colloids into liquid crystal is dominated by the elastic energy $F = \frac{1}{2}\int \left\{K_1(\nabla \cdot \bm{n})^2 + K_2(\bm{n}\cdot\nabla\times\bm{n})^2 + K_3(\bm{n}\times\nabla\times\bm{n})^2\right\}d^3r$, where $K_1$, $K_2$ and $K_3$ are splay, twist and bend elastic constants, respectively, although the total energy additionally includes the surface energy due to finite surface anchoring of $\bm{n}(\bm{r})$ at the particle surfaces, the contribution of flexoelectric terms, and the energy of defect cores that can be treated as having a reduced order parameter or a biaxial nature[33]. The surfaces of handlebody colloids have regions with opposite curvature, thus inducing the corresponding distortions of $\bm{n}(\bm{r})$ that minimize elastic energy for perpendicular boundary conditions at their surface. This results in the appearance of additional self-compensating pairs of defects of opposite hedgehog charge, leading to a number of defects that exceeds the minimum number, $g - 1$, required by topology. In the experimentally studied systems, colloidal $g$-handlebodies typically induce $g + 1$ individual singularities. Of these, $g - 1$ defects are of the same charge and are dictated by the particle topology, and two additional defects with opposite signs (total hedgehog charge zero) appear to relax $\bm{n}(\bm{r})$ distortions to minimize the free energy. Because $K_2 < K_1 < K_3$ for pentyl cyanobiphenyl[18], some of the splay distortions in the holes of the handlebodies confined into thin cells transform into more complex configurations, as demonstrated by spiralling dark and bright brushes in PM and 3PEF-PM images (Fig. 1a–d, g and Supplementary Figs 3 and 8). Although the structures shown in Fig. 3g, h and Supplementary Fig. 11 are unstable because of their high free energy and are found to relax to other topologically equivalent stable and metastable configurations around individual colloidal handlebodies (Figs 1 and 2), there is a possibility that they could be stabilized by confinement in twisted liquid crystal cells, as previously observed for the "bubble gum" configurations formed around colloidal dimers[22].

We have characterized the Brownian motion of colloidal handlebodies (Fig. 4a–c). Their diffusion in a planar cell with thickness much larger than the diameter of the handlebody is highly anisotropic (Fig. 4a, d) and easier along $\bm{n}_0$ than perpendicular to it. The slopes of mean square particle displacements (MSDs), shown in Fig. 4a for a solid torus ($g = 1$), yield diffusion coefficients $D_x = 0.0023$ µm$^2$ s$^{-1}$ and $D_y = 0.0034$ µm$^2$ s$^{-1}$ measured normal and parallel to $\bm{n}_0$, respectively. Being oriented with respect to $\bm{n}_0$ (Fig. 4d), particles also experience angular thermal fluctuations (Fig. 4b) with $\langle\Delta\theta^2\rangle$ of angular displacements (MSD$_\theta$) initially increasing linearly with the lag time $\tau$ and then saturating as a result of the elasticity-mediated alignment. The width of the histogram distribution of the angle $\theta$ between the axis of revolution of the solid torus and $\bm{n}_0$ (Fig. 4b, inset) is $9.6\times10^{-3}$ rad. The lateral diffusion of these colloids along directions perpendicular to $\bm{n}_0$ in homeotropic nematic cells is isotropic (Fig. 4c). However, when characterized in the particle's body frame, diffusion of

$g > 1$ handlebodies is anisotropic because of their shape. For example, $g = 2$ particles diffuse more easily along their long axis $a$ (Fig. 4c) crossing the centres of the two holes than along the short axis $b \perp a$ (ref. 27). The average diffusion of $g$-handlebodies having the same diameter of rings decreases with increasing $g$ (Fig. 4c). While being elastically trapped in the vicinity of the handlebodies, defects accompanying the particles also undergo thermal fluctuations.

In addition to laser tweezing and local melting, the relation between defects in $n(r)$ and $\chi$ can also be probed by applying an electric field $E$ (Fig. 4d) that causes the rotation of $n$ towards $E$ as a result of the liquid crystal's positive dielectric anisotropy. Two types of response have been observed. When $E$ is increased continuously, colloidal handlebodies reorient while preserving their alignment with respect to the director and following its reorientation towards $E$ normal to the substrates (Fig. 4e). However, because of slow rotation of the ring compared with the roughly 10-ms response time of $n(r)$, an abrupt application of $E$ simply alters $n(r)$ around the particle while preserving the initial particle alignment in the cell. For a solid torus ($g = 1$), this causes the original $n(r)$ to transform into a topologically equivalent configuration with two disclination loops (Fig. 4f, g). Using different voltage-driving schemes, colloidal handlebodies and structures around them can be switched between the two bistable orientations and $n(r)$ configurations shown in Fig 4d, f that are stable at no applied field. All observed transformations of $n(r)$ and orientations of colloids are again found to satisfy the relation $\Sigma_i m_i = \pm\chi/2$.

Our study experimentally supports the procedure for assignment of signs of topological defects in three-dimensional $n(r)$ textures before their summation that requires a global point of reference, a "base point", that serves as a global choice for the overall sign of $n(r)$, held fixed during any smooth deformation of the director complexion. Once fixed, the non-polar director field can be decorated with a vector, and the use of vector-field lines allows us to assign unambiguous signs to the defects (Fig. 3a–f)[32]. Although used for many decades[10], the convention that all hyperbolic point defects and disclination loops of −1/2 strength have charge "−1" whereas all radial defects and disclination loops of +1/2 strength have hedgehog charge "+1" fails to properly describe the topological charge conservation of hedgehog charges in the studied three-dimensional textures. The base point and the use of vector field lines in the liquid crystal texture until the signs of the hedgehog charges are assigned with respect to this base point allow a proper addition of hedgehog charges to the net charge of $\pm\chi/2$. The signs of topological point defects and the entire $n(r)$ structure induced by handlebody colloids then depend on the direction of vector field lines at the base point and can be reversed by flipping this direction to an opposite one, because of the non-polar nature of nematic liquid crystals[33]. It is only through the use of a base point that defects in different places can be added together like charges so that the net topological charge is conserved. Our approach also describes topological charge conservation in liquid-crystal textures studied previously, as we show in the Supplementary Information with an example of a colloidal dimer and surrounding $n(r)$. One can assign and add hedgehog charges in $n(r)$ of samples with multiple colloidal particles having the same or different $\chi$, and the addition of each separate particle always contributes a net $\pm\chi/2 = -m_c$ to the topological charge distribution of particle-induced bulk defects compensating for $m_c$ and ensuring charge conservation. However, charges induced by particles can have opposite signs even within the same texture, thus enabling charge annihilation in the textures surrounding these particles, as we show in the Supplementary Information with an example of colloidal dimers.

We have designed and fabricated topologically distinct handlebody-shaped colloidal particles and explored the interplay between the topology of colloids and the defects that they induce in a uniformly aligned liquid crystal. These handlebody colloids are accompanied by topological defects with the net hedgehog charge always equal to half of the Euler characteristic of the particle surface. Topological colloids and the established procedure for the assignment and summation of topological charges in liquid crystals will enable basic studies of topological manifolds and the interplay between particle topology and order and disorder[7,33] with these model systems. Beyond the exploration of the topology of colloids, fields and defects, the experimental arena we have developed may enable the design of topology-dictated elastic colloidal interactions and reconfigurable self-assembly in liquid crystals[18], the entrapment and scaffolding of nanoparticles by particle-induced defects[23], the self-assembly of reconfigurable topological memory devices[14], and electrooptic and photonic devices based on bistable switching between different states with distinct director configurations and orientations of particles.

**METHODS SUMMARY**

Fabrication of silica (SiO$_2$) particles with handlebody topology involved the following procedure. First, a 90-nm sacrificial layer of aluminium was sputtered on a silicon wafer. Next, a 1-µm silica layer was deposited on the aluminium by plasma-enhanced chemical vapour deposition. Photoresist AZ5214 (Clariant AG) was spin-coated on the silica layer. The pattern of rings was defined in the photoresist by illumination at 405 nm with a direct laser-writing system (DWL 66FS; Heidelberg Instruments) and then in the silica layer by inductively coupled plasma etching. Finally, the photoresist was removed with acetone and the aluminium was wet-etched with sodium hydroxide aqueous solution so that the handlebody particles were released and then re-dispersed in deionized water (Supplementary Fig. 1). To define perpendicular boundary conditions for $n(r)$ on the surface of particles, they were treated with an aqueous solution (0.05 wt %) of N,N-dimethyl-N-octadecyl-3-aminopropyl-trimethoxysilyl chloride (DMOAP) and then re-dispersed in methanol. After the addition of pentyl cyanobiphenyl and the evaporation of methanol at 70°C overnight, the ensuing nematic dispersion was infiltrated into cells composed of indium–tin-oxide (ITO)-coated glass plates separated by glass spacers defining the cell gap. Cell substrates were treated with DMOAP to achieve perpendicular $n_0$ or coated with polyimide PI2555 (HD Microsystems) for in-plane alignment of $n_0$ defined by rubbing. Optical manipulation and three-dimensional imaging of samples were performed with an integrated setup of holographic optical tweezers and 3PEF-PM (Supplementary Fig. 2)[23,30] built around an inverted microscope IX 81 (Olympus) and using a 100× oil-immersion objective (numerical aperture 1.4). Holographic optical tweezers used a phase-only spatial light modulator (Boulder Nonlinear Systems) and an ytterbium-doped fibre laser (IPGPhotonics) operating at 1,064 nm. 3PEF-PM employed a tunable (680–1,080 nm) Ti–sapphire oscillator (Coherent) emitting 140-fs pulses at a repetition rate of 80MHz, and a photomultiplier tube detector H5784-20 (Hamamatsu)[30].

**Acknowledgements.**
We thank P. Chen, N. Clark, J.-i. Fukuda and S. Žumer for discussions. This work was supported by the International Institute for Complex Adaptive Matter and the National Science Foundation grants DMR-0844115 (Q.L., S.H. and I.I.S.), DMR-0820579 (B.S. and I.I.S.), DMR-0847782 (B.S., Q.L. and I.I.S.), PHY11-25915 (R.D.K., R.B.K., T.C.L. and I.I.S.) and DMR-1120901 (R.D.K. and T.C.L.). R.B.K., R.D.K., T.C.L. and I.I.S. thank the Kavli Institute for Theoretical Physics for their hospitality while this work was being discussed and prepared for publication.

**Author Contributions.** B.S., Q.L. and I.I.S. performed experimental work. Q.L., S.H. and I.I.S. designed and fabricated particles. B.S. and I.I.S. reconstructed director fields induced by colloids. T.C.L. and I.I.S. characterized topological charges of defects in particle-induced director fields. R.B.K., R.D.K., T.C.L. and I.I.S. proposed models of field transformations satisfying topological constraints and explained the relations between genus of colloids and the net topological charge of liquid crystal defects. I.I.S. conceived the project, designed experiments, provided funding and wrote the manuscript. All authors edited and commented on the manuscript.

**Figures**

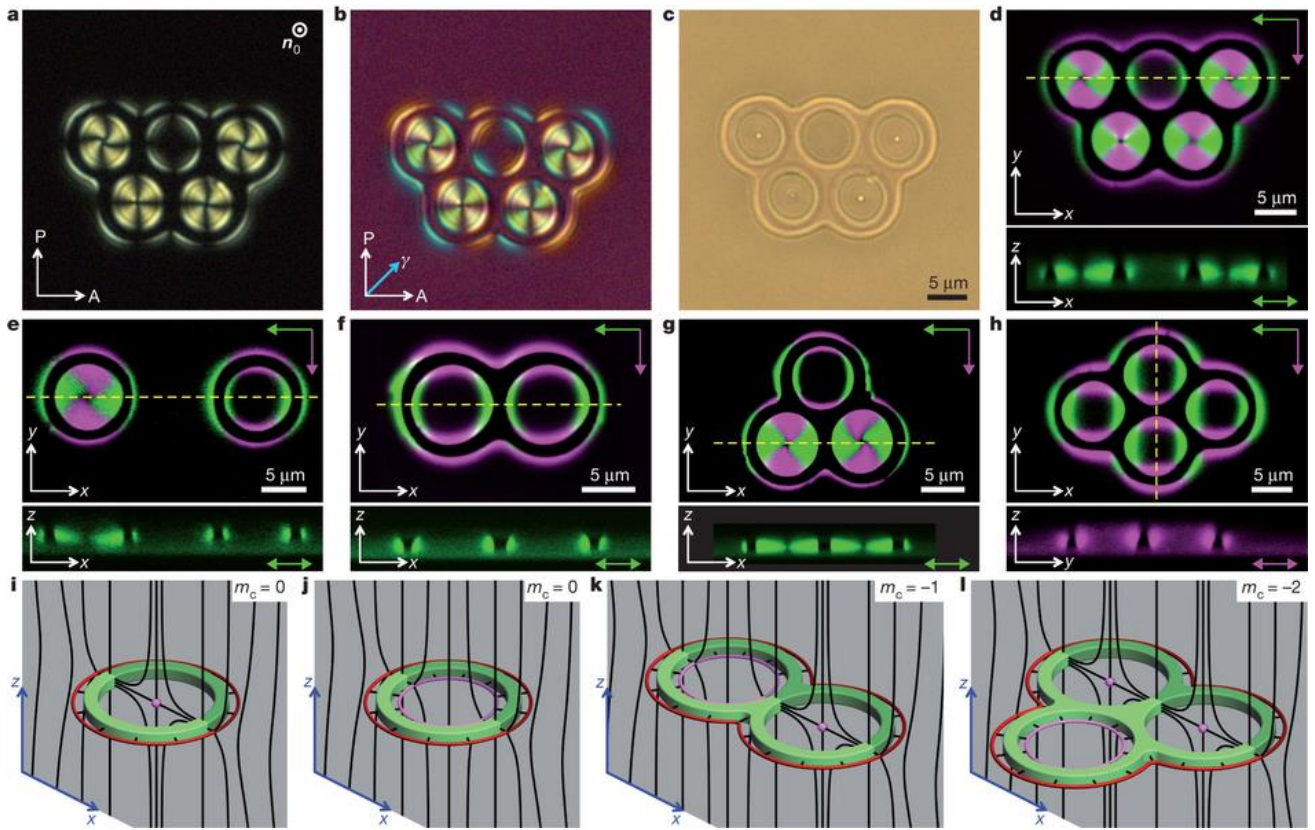

**Figure 1 | Colloidal handlebodies aligned orthogonal to the far-field director.** a–d, $g = 5$ handlebodies and induced $n(r)$ structures imaged by PM without (a) and with (b) a retardation plate, bright-field microscopy (c) and 3PEF-PM(d) techniques. e–h, 3PEF-PM textures of single (e), double (f), triple (g) and quadruple (h) handlebodies. P, A and $\gamma$ mark the crossed polarizer, analyser and a slow axis of a retardation plate (aligned at 45° to P and A), respectively. In the images (d–h) obtained by overlaying 3PEF-PM fluorescence intensity patterns for two orthogonal polarizations of excitation light, green and magenta colours correspond to the polarization directions marked by green and magenta arrows, respectively. Cross-sectional $xz$ and $yz$ images were obtained along yellow lines shown on the corresponding in-plane images. i–l, Diagrams of $n(r)$ (black lines) around $g$ handlebodies. Red and magenta lines show outer and inner disclination loops of $m = -1$ and $m = +1$ hedgehog charges, respectively. Magenta spheres show $m = +1$ hyperbolic point defects.

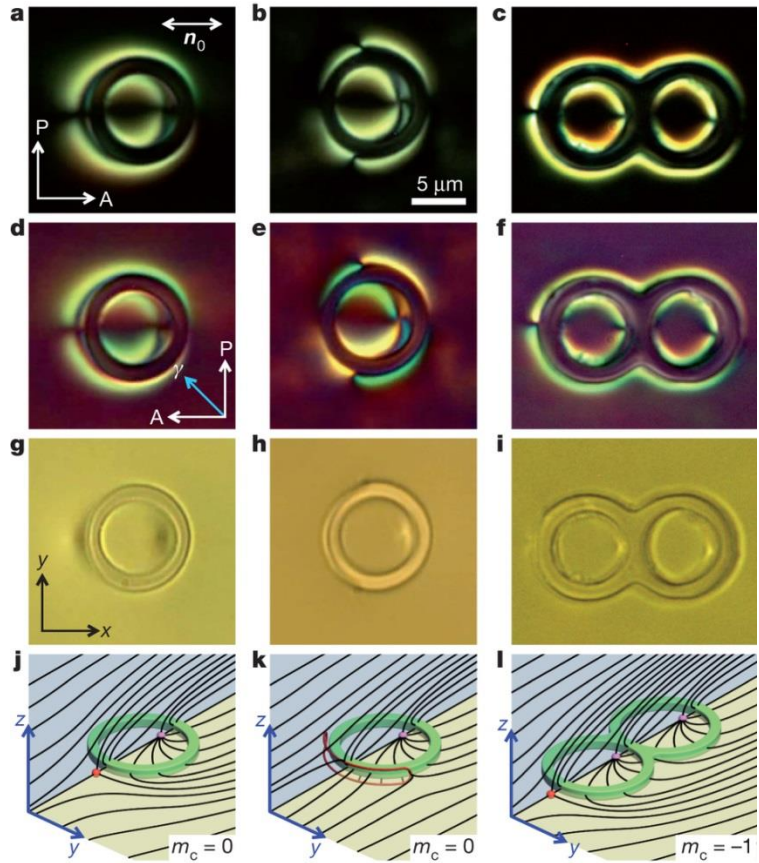

**Figure 2 | Colloidal g handlebodies aligned parallel to the far-field director.** a–i, Polarizing (a–f) and bright-field (g–i) textures and corresponding diagrams of $n(r)$ for different colloidal tori. Magenta and red spheres show the $m = +1$ and $m = -1$ hyperbolic point defects, respectively. j–l, The red loop in k shows a curved half-integer disclination ring with hedgehog charge $m = -1$ observed when a hyperbolic point defect near a solid torus (a, j) opens into a disclination loop (b, k). The black lines in j–l depict $n(r)$ in the plane of colloidal handlebodies (yellow) and in the plane orthogonal to the handlebodies (blue), both planes being parallel to $n_0$.

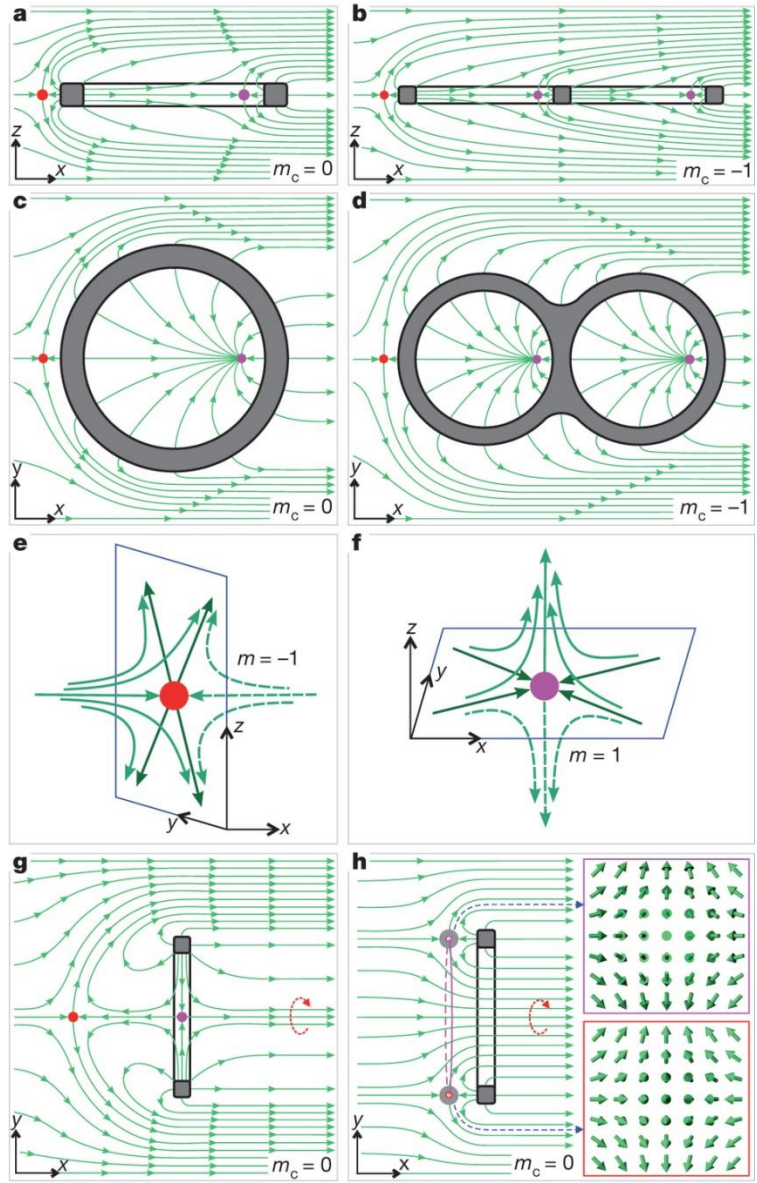

**Figure 3 | Director field and point defects around colloidal handlebodies.** a–d, Diagrams showing the vector-field representations of $n(r)$ (green lines with arrows) around single (a, c) and double (b, d) colloidal handlebodies in the plane of the rings (c, d) and in planes orthogonal to them (a, b). e, f, Diagrams of $n(r)$ around hyperbolic point defects of negative (e) and positive (f) topological signs shown by red and magenta filled circles. g, h, Diagrams showing elastic energy-costly unstable $n(r)$ structures with point defects of opposite $m = \pm 1$ (g) hedgehog charges and non-singular twist-escaped configuration (h) near a handlebody oriented perpendicular to $n_0$. The insets in h show the detailed vector field of the escaped axially symmetric configuration corresponding to non-singular $n(r)$ of $m = 0$. The cross-section of the field around a handlebody resembles that of an integer-strength disclination loop and is compensated for by an integer-strength disclination loop of opposite strength; the singularity of the latter is removed by an "escape in the third dimension" by means of continuous deformations as shown in the insets to h.

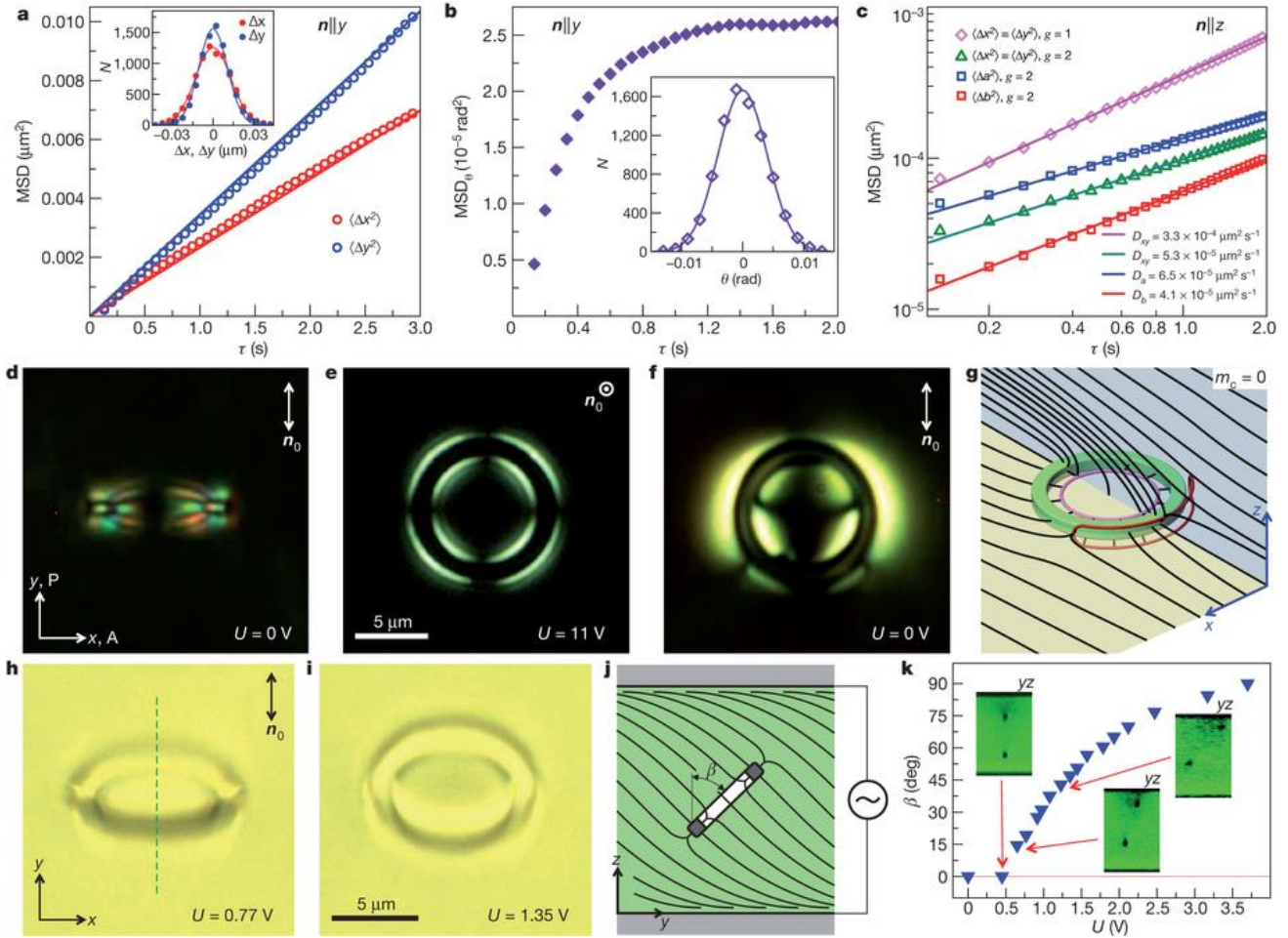

**Figure 4 | Diffusion and electric control of colloidal handlebodies in a nematic liquid crystal.** a, MSD of a solid torus parallel (blue circles) and perpendicular (red circles) to $n_0$ versus $\tau$ in a planar nematic cell of thickness $d = 16$ μm. The inset shows the corresponding experimental displacement histograms (red and blue filled circles) at $\tau = 0.067$ s, and Gaussian fits (red and blue lines). b, Plot of $MSD_\theta$ of a colloidal torus against $\tau$ and a histogram of angular orientations with respect to $n_0$ (inset). c, MSDs of $g = 1$ and $g = 2$ handlebodies in a homeotropic nematic cell of $d = 3$ μm along $x$, $y$, $a$ and $b$ axes; solid lines are linear fits to the data. Note the decreased mobility of handlebodies due to the thin cell confinement. d–f, PM textures of a planar nematic cell of $d = 17.5$ μm in the vicinity of a colloidal solid torus at no field (d), after continuous increase of applied alternating-current voltage to $U = 11$ V (1 kHz) within about 10 s (e), and after voltage $U = 14.8$ V was applied to the cell abruptly (as a square pulse modulated at 1 kHz) and then switched off (f). The image in f was taken at no applied field and after the torus and $n(r)$ had relaxed to the long-lived metastable state with the torus parallel to $n_0$. g, Diagram of $n(r)$ (black lines) in a texture shown in f. h, i, Bright-field images of a torus in a planar nematic cell of $d = 17.5$ μm reorienting with the liquid crystal director under the applied field $E$ normal to the image. j, Diagram showing $U$-controlled $n(r)$ deformations and rotation of the torus in the vertical $yz$ plane. k, Plot of torus tilt angle $\beta$ against $U$; the insets show 3PEF-PM cross-sectional images along the green dashed line marked in h at corresponding $U$.

# Supplementary Information

**I. Methods.** Microfabrication of colloidal $SiO_2$ (silica) *g*-handlebodies - or *g*-tori for short - involved the following procedure (Supplementary Fig. S1). First, a 90 nm thin layer of aluminum was sputtered on silicon wafer as a sacrificial layer. Second, a 1 μm $SiO_2$ layer was deposited on the aluminum layer using plasma-enhanced chemical vapor deposition. Then a thin layer of photoresist AZ5214 (Clariant AG) was spin-coated on the silica layer and the pattern of rings was defined by direct laser writing at 405 nm using a semiconductor laser of a direct laser writing system DWL 66FS (Heidelberg Instrument, Germany) followed by the inductively coupled plasma etching of the tori into the silica layer. Then the photoresist was removed by acetone (Supplementary Fig. S1). The aluminum layer was wet-etched by sodium hydroxide aqueous solution and the handlebody particles were released into the solution. They were washed out by deionized water (three times) to obtain an aqueous dispersion of colloidal *g*-handlebodies. To define the perpendicular surface boundary conditions for **n(r)** on the surface of colloids, they were treated by 0.05 wt% of N,N-dimethyl-N-octadecyl-3-aminopropyl-trimethoxysilyl chloride (DMOAP) aqueous solution and then re-dispersed into methanol. This dispersion was then mixed with 5CB to obtain a nematic dispersion after methanol evaporation overnight at an elevated temperature of 70 °C. The dispersion was infiltrated into cells comprised of two indium-tin-oxide (ITO) coated glass plates separated by glass spacers defining the cell gap. Cell substrates were treated with DMOAP to achieve perpendicular $\mathbf{n_0}$ or coated with polyimide PI2555 (HD Microsystems) for in-plane alignment of $\mathbf{n_0}$ defined by unidirectional rubbing.

Optical manipulation and 3D imaging was performed with a setup composed of holographic optical tweezers (HOT) and 3PEF-PM (Supplementary Fig. S2)[1,2] built around an inverted microscope IX 81 (Olympus). HOT utilized a reflective, phase-only spatial light modulator (SLM) obtained from Boulder Nonlinear Systems and an Ytterbium-doped fiber laser (YLR-10-1064, IPG Photonics) operating at 1064 nm. The SLM controlled the phase of the laser beam on a pixel-by-pixel basis according to the computer-generated holographic patterns at a refresh rate of 30 Hz. This phase-modulated beam was imaged at the back aperture of the microscope objective while creating the spatial trap intensity pattern in a sample. For 3PEF-PM imaging,[2] we have employed a tunable (680-1080 nm) Ti-Sapphire oscillator (Chameleon Ultra II, Coherent) emitting 140 fs pulses at a repetition rate of 80 MHz. The laser wavelength was tuned to 870 nm for the three-photon excitation of 5CB molecules. The 3PEF-PM signal was collected in epi-detection mode with a photomultiplier tube (H5784-20, Hamamatsu). An Olympus 100× oil-immersion objective with high numerical aperture of 1.4 was used for both imaging and optical trapping. A detailed description of the used experimental setups is described elsewhere.[1,2]

**II. Imaging and reconstruction of director structures around colloidal *g*-handlebodies in LCs.**
Colloidal *g*-handlebodies dispersed in a nematic LC were manipulated and imaged using the optical setup shown in the Supplementary Fig. S2. Supplementary Figs. S3-S7 show the textures of colloidal *g*-tori of genus *g* = 1, 2, ... , 5 and with homeotropic boundary conditions obtained in homeotropic LC cells using bright-field microscopy, PM, and 3PEF-PM.[1,2] All *g*-tori tend to spontaneously align perpendicular to $\mathbf{n_0}$ and are surrounded by single half-integer exterior disclination loops but induce different defects located within their interior. For example, PM textures (Supplementary Fig. S3a-d) show that the holes of *g*-tori contain either a singular half-integer disclination loop or a hyperbolic point defect in $\mathbf{n(r)}$. PM images with a full-wave retardation plate inserted between crossed polarizers and after the LC cell exhibit textures with colors directly related to the orientation of the spatially varying $\mathbf{n(r)}$ (which is also the optic axis) with respect to the "slow axis" γ. When the phase retardation due to the sample is small, as in the case of LC cells with $\mathbf{n(r)}$-distortions induced by particles in homeotropic cells, the blue color indicates that $\mathbf{n(r)}$ is at orientations close to parallel to γ, whereas yellow color corresponds to $\mathbf{n(r)} \perp γ$. Bright-field microscopy and PM mages (obtained both with and without the retardation plate) provide information about the location of defects with respect to colloids as well as a two-dimensional "view" of a complex three-dimensional director field around the *g*-tori (Figs. 1 and 2 and Supplementary Figs. S3-S8 and S10) as well as allow us to probe the dynamics of particle and director rotation due to photothermal melting (Fig. 2) and applied electric fields (Fig. 4).

    3PEF-PM textures (Supplementary Fig. S3e-p) allow one to reconstruct the details of complex three-dimensional director fields around *g*-tori. The maximum-intensity areas in the fluorescence textures correspond to the linear polarization of the laser excitation light being parallel to $\mathbf{n(r)}$ and anisotropic LC molecules. The dark areas with minimum fluorescence are observed when polarization of excitation light is perpendicular to the LC director. The strong ($\propto \cos^6 α$) dependence of the 3PEF-PM fluorescence on the angle α between $\mathbf{n(r)}$ and the linear polarization of excitation light[1] allows for reconstruction of the director field around colloidal tori (Supplementary Figs. S3-S7) based on multiple images obtained for different cross-sectional planes and for different polarizations

of excitation laser light. Color-coded 3PEF-PM images for different mutually-orthogonal laser excitation polarization states (similar to the ones shown in the Supplementary Figs. S3-S7) were overlaid on top of each other to obtain the images shown in the main-text (Figs. 1 and 2). The intrinsic three-dimensional optical resolution of the nonlinear 3PEF-PM imaging due to the highly localized multiphoton excitation enables three-dimensional nondestructive imaging of the director fields around colloidal *g*-tori (Figs. 1, 2, 4 and Supplementary Figs. S3-S10).

**III. Multistable structures and effects of confinement.** In homeotropic nematic cells, *g*-tori with perpendicular boundary conditions tend to self-align perpendicular to $\mathbf{n_0}$. They are surrounded by single half-integer exterior disclination loops, but different types of defects occur within their holes (Supplementary Fig. S8a-d). Two stable defect configurations are commonly observed inside the *g*-tori holes: a hyperbolic hedgehog point defect (Supplementary Fig. S8a, b, e, g, h) and a half-integer disclination loop (Supplementary Fig. S8c, d, f, i). Both defects were observed in the colloidal tori of different ring diameters ranging from 5 to 10 μm (Supplementary Fig. S3). The point defects exhibit Brownian motion[3] but their diffusion is elastically confined to the center of the *g*-torus hole (Supplementary Fig. S8a, b). Local melting with subsequent quenching of the LC in the torus interior using optical tweezers often results in bistable switching between these two structures.

In addition to external field and laser beams, director structures around *g*-tori immersed in planar LC cells can be controlled by varying the cell thickness *d*. Supplementary Figure S10 shows the 3PEF-PM textures and schematic of the director field distortions around the torus in the planar cell with the thickness much larger than the torus diameter. In this geometry, the torus aligns normal to $\mathbf{n_0}$ and has two half-integer disclination loops (Supplementary Fig. S9c, d) around it, as in the case of the homeotropic cell (Supplementary Figs. S3 and S8). In the planar cells with *d* smaller than the torus diameter *D*, the confinement causes tilting of the torus with respect to $\mathbf{n_0}$ (Supplementary Fig. S10) and the tilt angle varies with the ratio *d/D*. Although $\mathbf{n(r)}$ becomes more distorted (Supplementary Fig. S10a, d, e), the total topological hedgehog charge of all disclination loops and point defects remains conserved and $\sum_i m_i = \pm \chi/2$.

In addition to the experimentally observed $\mathbf{n(r)}$-configurations, a number of other structures (such as the ones shown in Fig. 3g, h and Supplementary Fig. S11) may be possibly realized under different experimental conditions, although their appearance is hindered by the high free energy of elastic distortions. Furthermore, many topologically similar textures can be obtained by varying positions of point defects and disclination loops (compare Fig. 3g and Supplementary Fig. S11). Despite of being unstable in the bulk of uniformly aligned nematic LCs, there is a possibility that these structures can be stabilized by means of cell confinement and twisted configurations of **n(r)** in the nematic LC hosts.

**IV. Director field around spherical particles and their dimers.** The described procedure of assignment and summation of hedgehog charges using a basepoint can be applied to LC-colloidal systems that were studied previously. For example, the Supplementary Fig. S12a, b shows PM textures of the elastic dipole[4] formed by a solid spherical colloidal particle immersed into a planar nematic LC cell. The total topological charge of defects in the ensuing **n(r)** around the colloid (Supplementary Fig. S12c, d) obeys the relation $\sum_i m_i = -m_c = \pm \chi/2$ as well, where $m = \pm 1$ is the topological charge of the point defect hedgehog (filled circles shown in Supplementary Fig. S12c, d) and $\chi = 2$ is the Euler characteristic of the sphere. Another example (which goes beyond a single-particle situation that we discussed previously) is a colloidal dimer formed by two microspheres with perpendicular boundary conditions and a twist-escaped integer-strength disclination loop (dubbed "bubble gum" in the previous studies).[5,6] The corresponding PM textures are shown in the Supplementary Fig. S12e, f. Our procedure of assignment of defect charges again requires the use of a basepoint, e.g. one of the two spherical particles for which we can choose vector field lines pointing outward and perpendicular to the particle's surface (Supplementary Fig. S12h). By assuring a continuous vector field around these particles, one can see that the hedgehog charges in **n(r)** around the two colloids have opposite signs. This can be shown by mapping **n(r)** decorated with a vector around them onto the order parameter space (a two-sphere for the vector field), although both of them have elementary charge with the absolute value of unity, e.g. $m_c = \pm\chi/2 = \pm 1$. Mapping the texture of the looped twist-escaped disclination onto the corresponding two-sphere order parameter space yields its net hedgehog charge equal to zero. Since the two colloids induce opposite topological hedgehog charges, this again assures that the total topological charge is equal to zero (conserved), regardless of the choice of the global reference point and vector field direction that only determines which of the two colloidal spheres is inducing "−1" and which is inducing "+1" hedgehog charge in the field. Since **n(r)** is nonpolar, one obtains an equally appropriate description of this texture upon consistently changing the signs of all hedgehog charges to the opposite ones. The example of a colloidal dimer with the "bubble gum" structure shows that topological charges induced by identical colloidal particles can have opposite signs (similar to, say, radial hedgehog defects in nematic textures being able to carry charges of opposite signs in the same nematic sample). It also shows that the hedgehog charges due to multiple particles in the same sample can annihilate each other while forming configurations of net zero charge, such as that of the twist-escaped disclination loop (Supplementary Fig. S12h) that embeds the particles into a sample with the uniform far-field director.

In the past studies of LC director fields, the assignment of signs of hedgehog charges due to point defects and disclination loops utilized a different convention,[7] which cannot describe topological charge conservation in our experiments but was surprisingly successful in describing some of the nematic textures in the past. According to this traditionally used convention, the hyperbolic point defects and disclination loops of strength $s = -1/2$ were commonly assigned topological charge of "–1" while radial point defects and $s = 1/2$ disclination loops were commonly assigned a hedgehog charge "+1". Following this convention, particle with vertical boundary conditions in the elastic dipole can be assigned a charge "+1" while the hyperbolic point defect next to it would always have charge "–1" (Supplementary Fig. S12c).[7] In the case of colloidal dimer (Supplementary Fig. S12e, f), a loop of a twist-escaped integer-strength disclination of strength $s = -1$ would need to be assigned a hedgehog charge of "–2", compensating the two "+1" hedgehog charges due to boundary conditions for the director around spherical particles (Supplementary Fig. S12g).[6] These charges can be also obtained by mapping the director field lines with nonpolar symmetry onto the order parameter space ($\mathbb{S}^2/\mathbb{Z}_2 \cong \mathbb{R}P^2$, the two-sphere with pairs of diametrically opposite points being identical). Although this "old" convention provides one way of understanding of the topological charge conservation for certain structures, it clearly fails to properly describe the hedgehog charges in nematic textures induced by handlebody particles studied here (see Figs. 1-4 in the main text). Using three-dimensional non-contact manipulation and imaging of director fields induced by handlebody colloids, we unambiguously demonstrate that this "old" convention is incapable of the proper assignment and summation of charges due to defects and textures in nematic LCs and thus the new charge assignment and summation procedure discussed here should be used instead.

**Supplementary References**

## Supplementary Figures

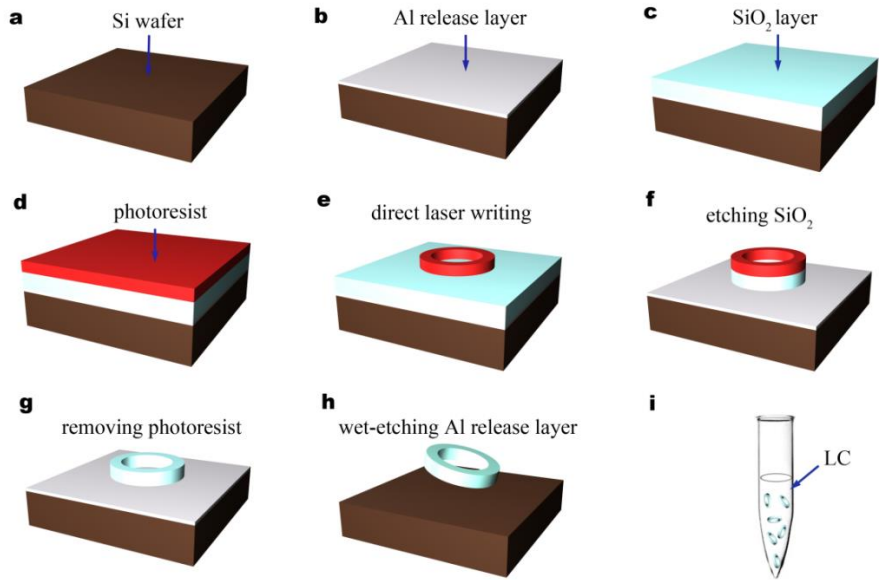

**Figure S1.** Preparation of $SiO_2$ colloidal tori dispersed in 5CB. (a,b) We sputter an aluminum release layer on a silicon wafer and then (c) deposit an $SiO_2$ layer and (d) spin-coat the photoresist. We then (e) expose patterns of the tori by direct laser writing, (f) etch $SiO_2$, (g) remove the photoresist, and (h) wet-etch the aluminum layer to release the colloidal $SiO_2$ tori. (i) The particles are then re-dispersed into a nematic LC after washing the particles in organic solvents and water as well as using solvent exchange and evaporation.

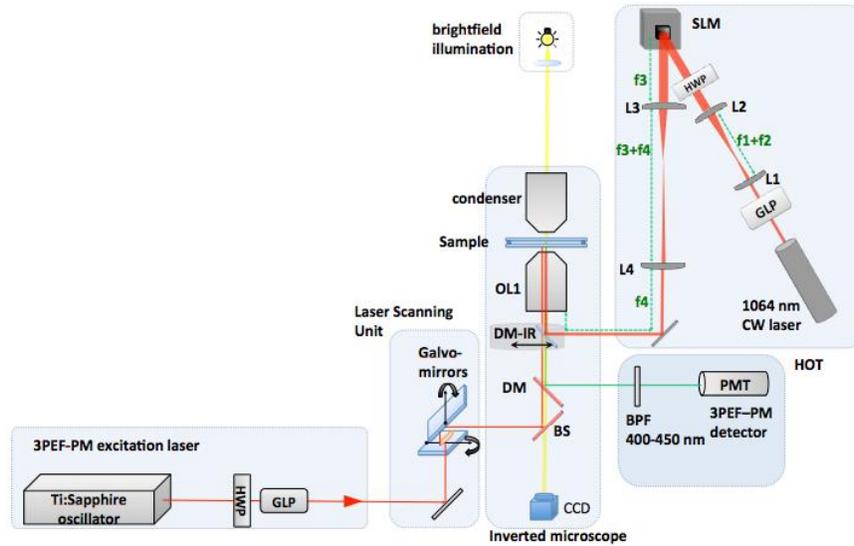

**Figure S2**. A schematic of the integrated 3D optical manipulation and imaging setup. HWP: half-waveplate, GLP: Glan laser polarizer, DM: dichroic mirror, BPF: bandpass filter, OL: objective lens, SLM: spatial light modulator, L1-L4: plano-convex lenses. f1-f4: focal length of the lenses, PMT: photomultiplier tube, BS: beam splitter.

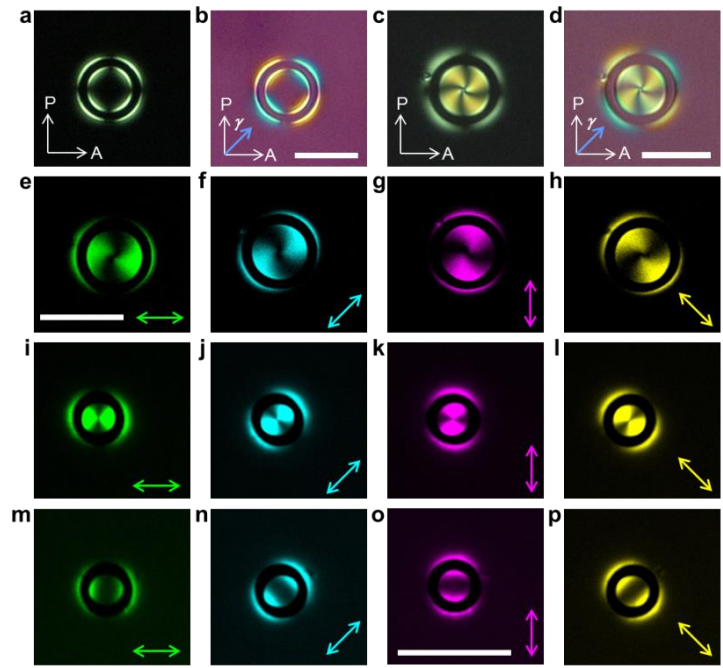

**Figure S3.** Single toroidal particles of different size and with perpendicular surface anchoring in homeotropic LC cells. (a-d) Polarizing microscopy textures of particles with an outer disclination loop and either an inner disclination loop (a, b) or a point defect (c, d). (e-p) 3PEF-PM textures of toroidal particles of different size and director configuration. Color corresponds to the fluorescence excited by light polarized in the direction shown by color double arrows. "P", "A" and "$\gamma$" show crossed polarizer, analyzer and a slow axis of a retardation plate, respectively. White scale bar is 10 µm.

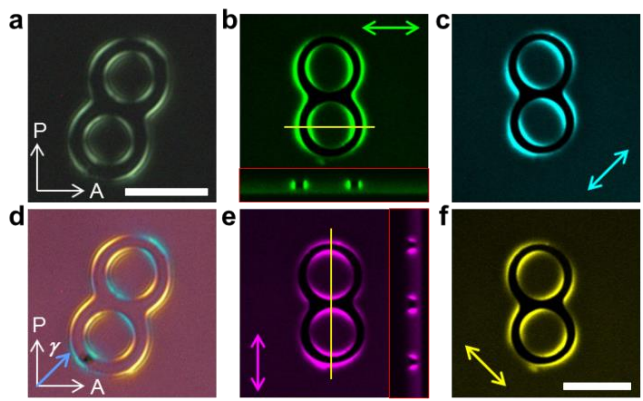

**Figure S4.** Polarizing (a, d) and 3PEF-PM (b, c, e, f) textures of colloidal double tori. Insets delineated by a red square show cross-sectional textures obtained along yellow lines. White scale bars are 10 μm.

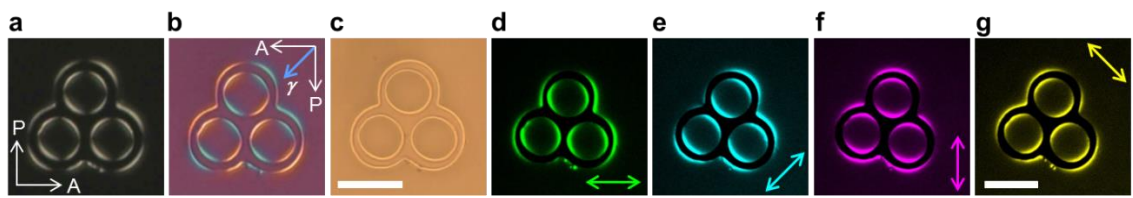

**Figure S5.** Polarizing (a-c) and 3PEF-PM (d-g) textures of colloidal triple tori with homeotropic anchoring in a homeotropic LC cell. White scale bar is 10 μm.

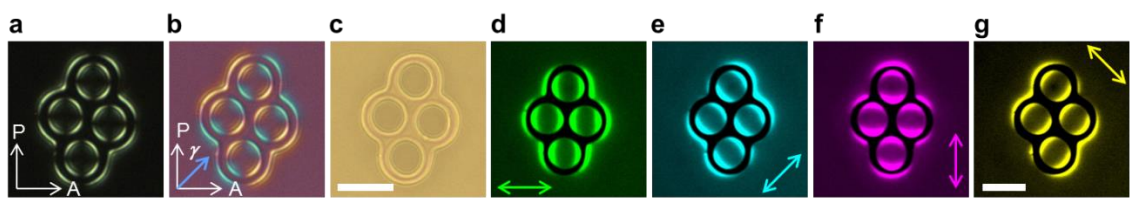

**Figure S6.** Polarizing (a-c) and 3PEF-PM (d-g) textures of colloidal quadruple tori with perpendicular surface anchoring in the homeotropic LC cell. White scale bars are 10 μm.

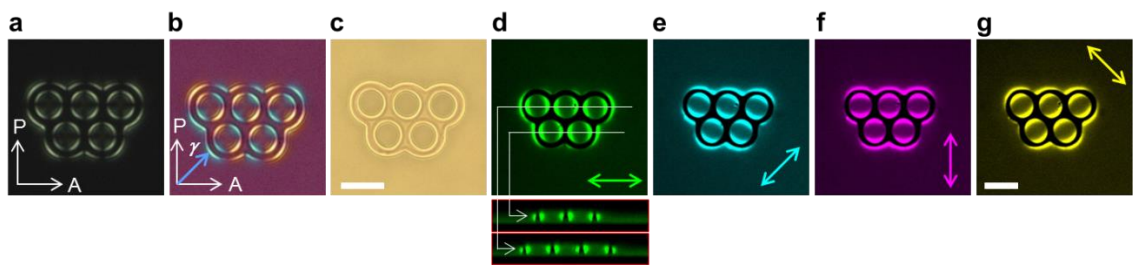

**Figure S7.** Polarizing (a-c) and 3PEF-PM (d-g) textures of colloidal quintuple tori with perpendicular surface anchoring in a homeotropic LC cell. Insets show the cross-sectional views of particles along the white lines. White scale bars are 10 μm.

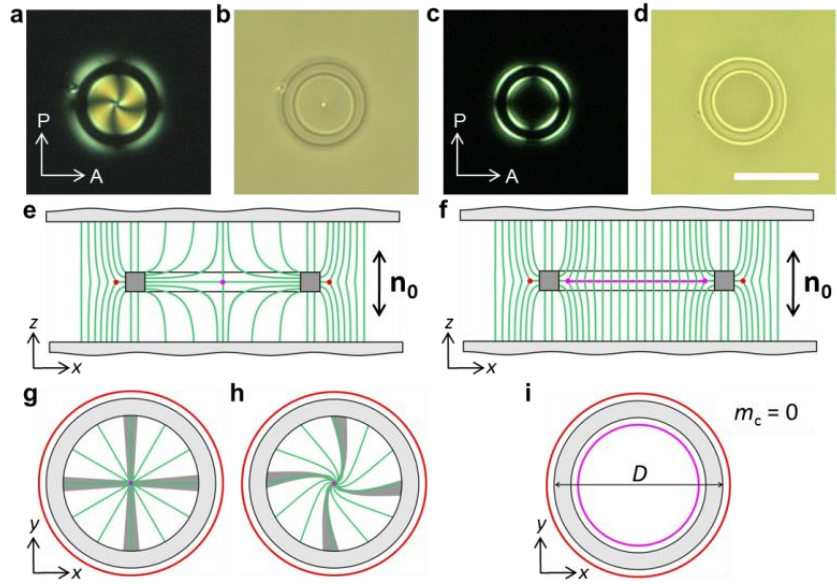

**Figure S8.** Director structures with point and ring defects induced by torus-shaped particles in a nematic LC. (a-d) Polarizing (a, c) and bright-field (b, d) optical micrographs of a single toroidal particle with homeotropic anchoring placed in the homeotropic liquid crystal cell, and (e-i) schematics of corresponding director field (green lines) and point (red and magenta filled circles) and line (red and magenta lines) defects. The white scale bar is 10 μm.

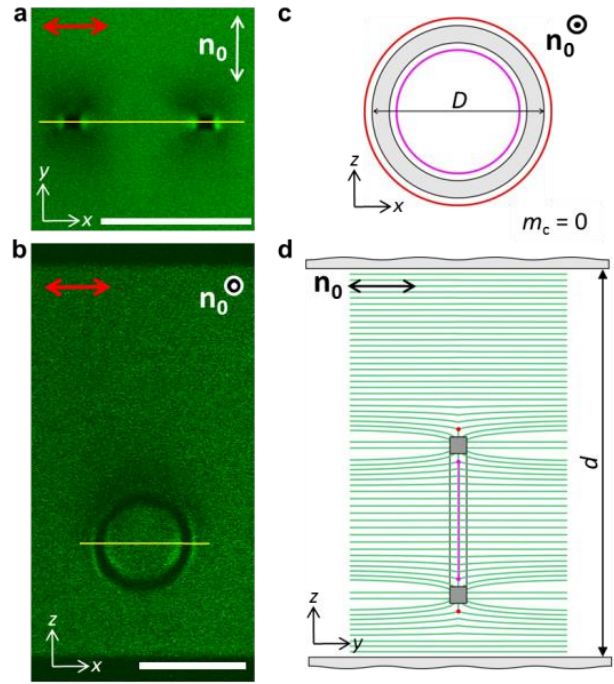

**Figure S9.** A single toroidal particle with perpendicular surface anchoring in a thick planar LC cell. (a, b) 3PEF-PM images of the colloidal torus and $\mathbf{n}(\mathbf{r})$ around it. (c, d) Schematics show the corresponding $\mathbf{n}(\mathbf{r})$ and line defects loops. Red double arrow shows the polarization direction of excitation light. White scale bars are 10 μm.

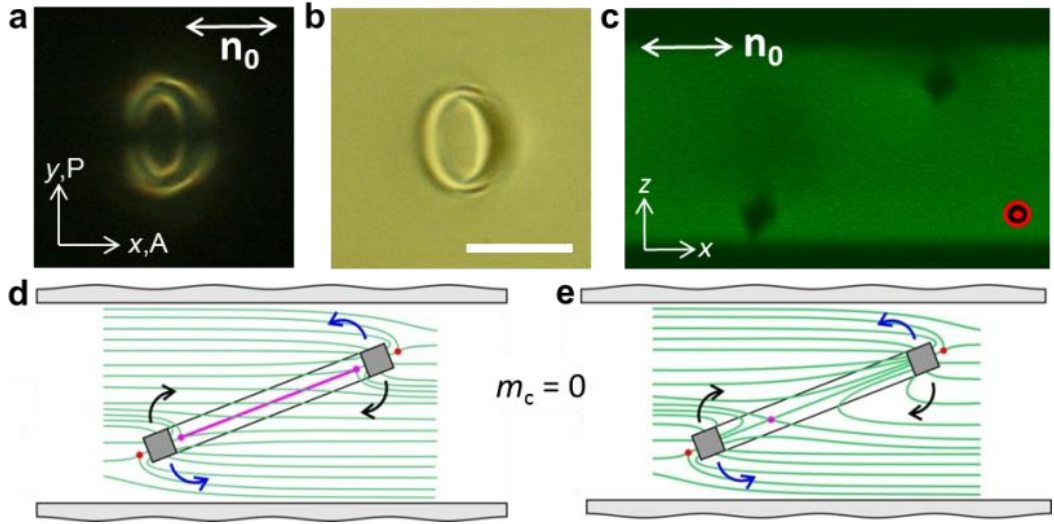

**Figure S10.** A toroidal particle with perpendicular surface anchoring in a thin (thickness is smaller than the diameter of the toroidal particle) planar LC cell. (a) Polarizing, (b) bright-field and (c) 3PEF-PM cross-sectional textures of the colloidal torus tilted between substrates and **n(r)** around it. (d, e) Schematics of corresponding **n(r)** and defects. The white scale bar is 10 µm.

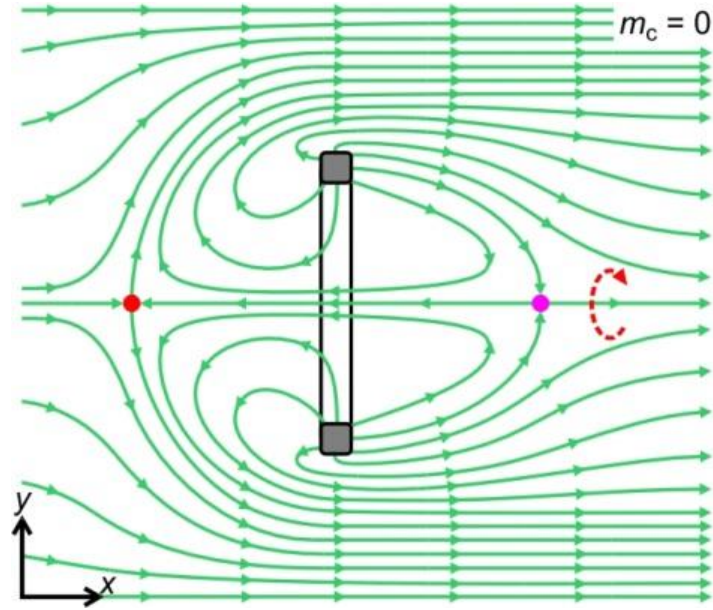

**Figure S11.** A schematic showing elastic-energy-costly unstable **n(r)**-structure with point defects of opposite hedgehog charges $m = \pm 1$ near a handlebody oriented perpendicular to $\mathbf{n_0}$ having $m_c=0$. Red and magenta filled circles show negative and positive charges of point defects, respectively.

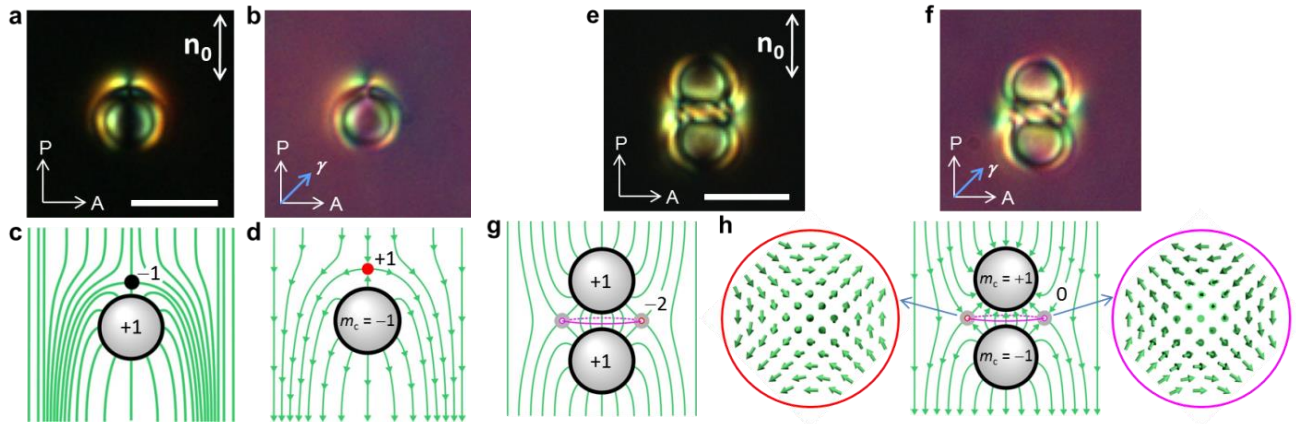

**Figure S12.** Charge summation in nematic LC samples with (a-d) single and (e-h) two spherical colloids. (a, b) Polarizing textures of an elastic dipole formed by spherical colloidal particles with perpendicular surface anchoring in a planar nematic LC cell and (c, d) schematics showing the corresponding **n(r)** (green lines) and hyperbolic point defect (black and red filled circles). (e, f) Polarizing textures of a "bubble-gum" dimer. (g) A schematic of **n(r)** with a non-singular twist-escaped "bubble gum" configuration around a dimer with center-to-center separation along **n₀** that has topological hedgehog charges marked according to the "old" convention (e.g. particles inducing "+1" charges while the "bubble gum" inducing a charge of "-2"). (h) A schematic of **n(r)** around a colloidal dimer similar to that shown in (g) but now presented using vector field lines and having topological hedgehog charges marked according to the new charge assignment procedure (e.g. particles inducing opposite ±1 charges while the "bubble gum" having zero charge). The insets in (h) show the detailed vector-field of the escaped axially symmetric configuration corresponding to nonsingular **n(r)**. The white scale bars shown in (a, e) are 5 μm. The colloid's charge $m_c$ induced by particles in the vector field lines is marked on schematics (d, h). The schematics (c, g) have hedgehog charges of particles and defects marked as determined using the "old" convention.